\documentclass[%
 reprint,
 amsmath,amssymb,
 aps,nofootinbib
]{revtex4-1}
\usepackage{dsfont}
\usepackage{graphicx}
\usepackage{dcolumn}
\usepackage{bm}
\usepackage{color}
\usepackage{epsfig}
\usepackage{hyperref}
\usepackage{natbib}
\usepackage{float}
\usepackage{footmisc}
\usepackage{comment}
\definecolor{lightred}{rgb}{1,0.5,0.5}
\definecolor{lightgreen}{rgb}{0.5,1,0.5}
\definecolor{lightblue}{rgb}{0.5,0.5,1}
\definecolor{lightcyan}{rgb}{0.5,0.75,0.75}
\definecolor{lightmagenta}{rgb}{0.75,0.5,0.75}
\definecolor{customgreen}{rgb}{0.494,1,0.502}

\newcommand{\meV}{\mathinner{\mathrm{meV}}}

\newcommand{\keV}{\mathinner{\mathrm{keV}}}
\newcommand{\MeV}{\mathinner{\mathrm{MeV}}}
\newcommand{\GeV}{\mathinner{\mathrm{GeV}}}

\newcommand{\orcid}[1]{\href{https://orcid.org/#1}{\textcolor{black}{\,\includegraphics[height=2ex]{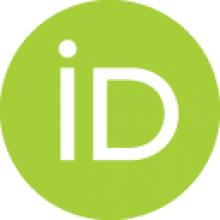}}}}

\begin{document}

\title{Thermal Perturbations from Cosmological Constant Relaxation}

\author{Lingyuan Ji\orcid{0000-0002-3947-7362}}
\affiliation{Department of Physics and Astronomy, Johns Hopkins University, 3400 N. Charles St., Baltimore, Maryland 21218, USA}

\author{David E. Kaplan\orcid{0000-0001-8175-4506}}
\affiliation{Department of Physics and Astronomy, Johns Hopkins University, 3400 N. Charles St., Baltimore, Maryland 21218, USA}

\author{Surjeet Rajendran\orcid{0000-0001-9915-3573}}
\affiliation{Department of Physics and Astronomy, Johns Hopkins University, 3400 N. Charles St., Baltimore, Maryland 21218, USA}

\author{Erwin H. Tanin\orcid{0000-0002-3204-0969}}
\affiliation{Department of Physics and Astronomy, Johns Hopkins University, 3400 N. Charles St., Baltimore, Maryland 21218, USA}

\begin{abstract}
We probe the cosmological consequences of a recently proposed class of solutions to the cosmological constant problem. In these models, the universe undergoes a long period of inflation followed by a contraction and a bounce that sets the stage for the hot big bang era. A requirement of any successful early universe model is that it must reproduce the observed scale-invariant density perturbations at CMB scales. While these class of models involve a long period of inflation, the inflationary Hubble scale during their observationally relevant stages is at or below the current Hubble scale, rendering the de Sitter fluctuations too weak to seed the CMB anisotropies. We show that sufficiently strong perturbations can still be sourced thermally if the relaxion field serving as the inflaton interacts with a thermal bath, which can be generated and maintained by the same interaction. We present a simple model where the relaxion field is derivatively ({\it i.e.} technically naturally) coupled to a non-abelian gauge sector, which gets excited tachyonically and subsequently thermalizes due to its nonlinear self-interactions. This model explains both the smallness of the cosmological constant and the amplitude of CMB anisotropies. 
\end{abstract}

\maketitle

\section{Introduction}
Cosmological observations have shown that the universe is presently undergoing accelerated expansion due to a form of energy density dubbed ``dark energy''. The inferred measured value of the dark energy density is $\sim 10 \text{ meV}^4$. This measurement constrains the value of the cosmological constant (CC) to be no bigger than $\sim 10 \text{ meV}^4$, an energy density that is at least 60 orders of magnitude smaller than known perturbative contributions to it from the Standard Model. The extreme fine tuning necessary to cancel these known contributions is known as the cosmological constant problem (see \cite{Weinberg:1988cp} for a review). A compelling possibility is to accept the expected large CC value, with all its radiative corrections included, as a starting point and have it compensated by a scalar field whose dynamics lowers its potential energy over time, gradually relaxing the effective CC in the process \cite{Abbott:1984qf}. Given an appropriate interaction structure and cosmology, the relaxation can be made to naturally stop when the CC is sufficiently fine-tuned, thereby naturally generating a tiny CC.

Refs.~\cite{Graham:2017hfr, Graham:2019bfu} put forward a simple, proof-of-principle model for this class of solution that not only accomplishes CC fine-tuning from a naturally large initial value, but also explains how this story could fit into the hot-big-bang epoch. In this scenario, the universe undergoes an expansion, a contraction, and an expansion again. During the first expansion, the universe is in an extremely long period of inflation. This period of inflation is produced by the rolling of a scalar field that has a small, but technically natural, slope. As the field rolls down its potential, the effective CC steadily decreases and eventually becomes slightly negative. At this time, the large contributions to the CC have been dynamically cancelled. But, we are left with a universe that is cold and empty due to the long period of inflation. However, the negative value of the CC halts the expansion of the universe and makes it contract. During this period of contraction, various small energy densities in the universe (such as the kinetic energy of the scalar itself) blueshift, producing a contracting hot universe with a small negative CC. The energy density produced in the contraction could conceivably excite UV degrees of freedom that could trigger a bounce, causing the universe to re-expand as a hot universe but with a small, dynamically relaxed CC. In this scenario, our current hot big bang epoch exists during this stage of re-expansion. 

The principal accomplishment of this class of models is that it allows for a factorized approach to solve the CC problem. Specifically, it permits  UV physics at very high energies to play a role in solving the CC problem. In these models, the tuning necessary to cancel the CC is accomplished in the infrared by a slowly rolling scalar field. The UV plays a role at high energies/temperatures where this dynamics can trigger a bounce causing the re-expansion of a contracting universe.

One important issue that has not been addressed in the cosmology of the minimal CC-relaxation model described above is the origin of the approximately scale-invariant spectrum of scalar perturbations observed in the cosmic microwave background (CMB). The minimal model has an inflationary period, and therefore scale invariance, built into it. However, in order to solve the CC problem the Hubble rate during its observationally-relevant last tens of $e$-folds must at most be as low as its value today, $H_0\sim 10^{-42}\GeV$. If the CMB anisotropies were to be seeded by the vacuum fluctuations during such an extremely low energy inflation, the resulting scalar power spectrum would be far too small. One may hope that the superhorizon evolution of the curvature perturbations in the contraction phase after inflation would lead to their growth, however under general assumptions superhorizon curvature perturbations are conserved in the absence of non-adiabatic pressure perturbations \cite{Wands:2000dp, Cardoso:2008gz}\footnote{In some gauges, e.g longitudinal gauge, there exist a curvature perturbation mode that is growing when the universe is contracting, however it has been shown, again under general assumptions, that this mode does not contribute to the observationally-relevant post-bounce constant mode \cite{Creminelli:2004jg, Bozza:2005wn, Bozza:2005qg, Cardoso:2008gz}.}. The latter is essentially because the homogeneous background $\rho$ and the large-scale perturbations $\delta \rho$ of the total energy density follow the same evolution equation, which leads to their ratio $\delta\rho/\rho$ being fixed. This suggests the need for a non-adiabatic pressure perturbation contribution or a different source of curvature perturbations.

In this paper, we improve on the the minimal CC-relaxation model, while keeping the broad-brush cosmological history described above unchanged, by coupling a field $\phi$ that relaxes the CC to a Yang-Mills sector axially. Such a derivative coupling generates a thermal bath of radiation which, in turn, damps the motion of $\phi$ without giving it a thermal mass \cite{Berghaus:2019whh}. The thermal fluctuations in the radiation fields cause fluctuations in $\phi$ that are potentially much stronger than those of quantum origin \cite{Bastero-Gil:2011rva}. With appropriate choices of parameters, this model simultaneously solves the cosmological constant problem and explains the observed CMB anisotropies. Moreover, the thermal damping on the rolling of $\phi$ improves considerably the degree of CC fine-tuning that can be achieved through relaxation. 

The rest of the paper is organized as follows. We present in Section~\ref{s:model} the particle physics content of our cosmological constant relaxation model as well as a brief summary of its cosmology. We then describe the background evolution of the universe through the key cosmological epochs in our model and the validity of the effective field theory under consideration in Section~\ref{s:background}. Next, we show in Section~\ref{s:perturbations} that our model can explain the density perturbations seen in the CMB, before concluding with Section~\ref{s:discussion}.

\section{Model \label{s:model}}
Our starting Lagrangian contains an axion-like field $\phi$ with a linear potential which couples axially to a dark, pure $SU(N_c)$ Yang-Mills sector
\begin{equation}
    \mathcal{L}=\frac{1}{2}\left(\partial\phi\right)^2+g^3\phi-\frac{1}{4}G^{a}_{\mu\nu}G^{a\mu\nu}-\frac{\alpha}{8\pi}\frac{\phi}{f_G}G^a_{\mu\nu}\tilde{G}^{a\mu\nu}\label{model}
\end{equation}
Here, $\alpha\equiv e_G^2/4\pi$ is the fine-structure constant of the gauge sector and $e_G$ (not $g$) denotes the associated coupling strength. The potential $V=-g^3\phi$ serves as the effective CC to be relaxed and we will sometimes refer to it as the CC. Any $\phi$-independent contribution to the CC are accounted for by appropriately shifting $\phi$ by a constant value.

\begin{figure}
    \centering
    \includegraphics[width=\linewidth]{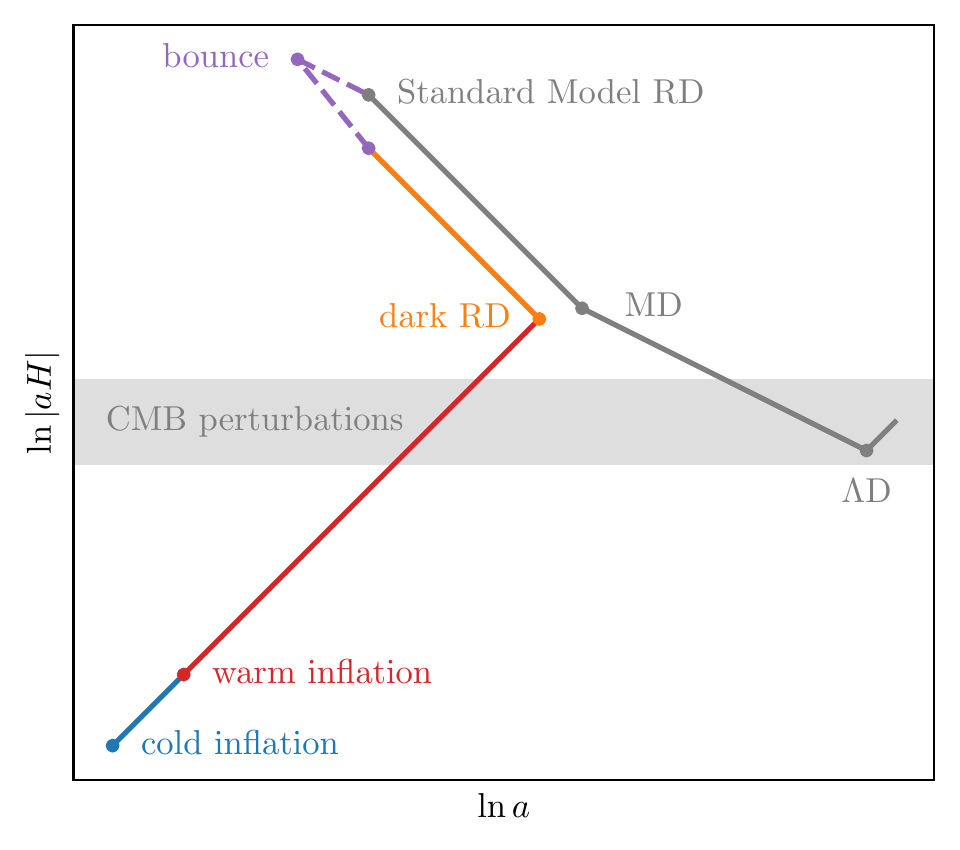}
    \caption{The scale factor v.s.\ comoving Hubble scale sketch (not-to-scale) of our model. While it is shown here that the scale factor today is greater than that at the end of warm inflation, the reverse is also possible.}
    \label{fig:history-sketch}
\end{figure}

To start with, the $\phi$ field is large and negative, with its large and positive potential $V=-g^3\phi$ driving a period of cold inflation, where the yet unoccupied gauge sector plays no role. The $\phi$ field proceeds to roll down its potential and, once it gains enough kinetic energy, the gauge fields become tachyonic, grow rapidly, and subsequently thermalize. Shortly after, the universe enters a period of warm inflation where the gauge-field thermal bath affects the background evolution through the thermal friction it exerts on $\phi$ and sources the dominant density perturbations from its thermal fluctuations. We assume the modes corresponding to the fluctuations seen in the CMB become superhorizon during this period. Warm inflation ends when $\phi$ has relaxed to a small negative value, which is followed by $\phi$ going through zero to positive values, making the potential $-g^3\phi$ negative and causing the universe to crunch. The now blue-shifting gauge-field radiation soon dominates the universe. Once it hits a high enough temperature, we assume that some UV dynamics are triggered causing the universe to bounce and re-expand, entering the standard hot big bang epoch. An example of such UV dynamics was discussed in \cite{Graham:2017hfr}, but many other such models may be possible.  A sketch of the full cosmological history is shown in Fig.~\ref{fig:history-sketch}.

\begin{figure}
    \centering
    \includegraphics[width=0.45\textwidth]{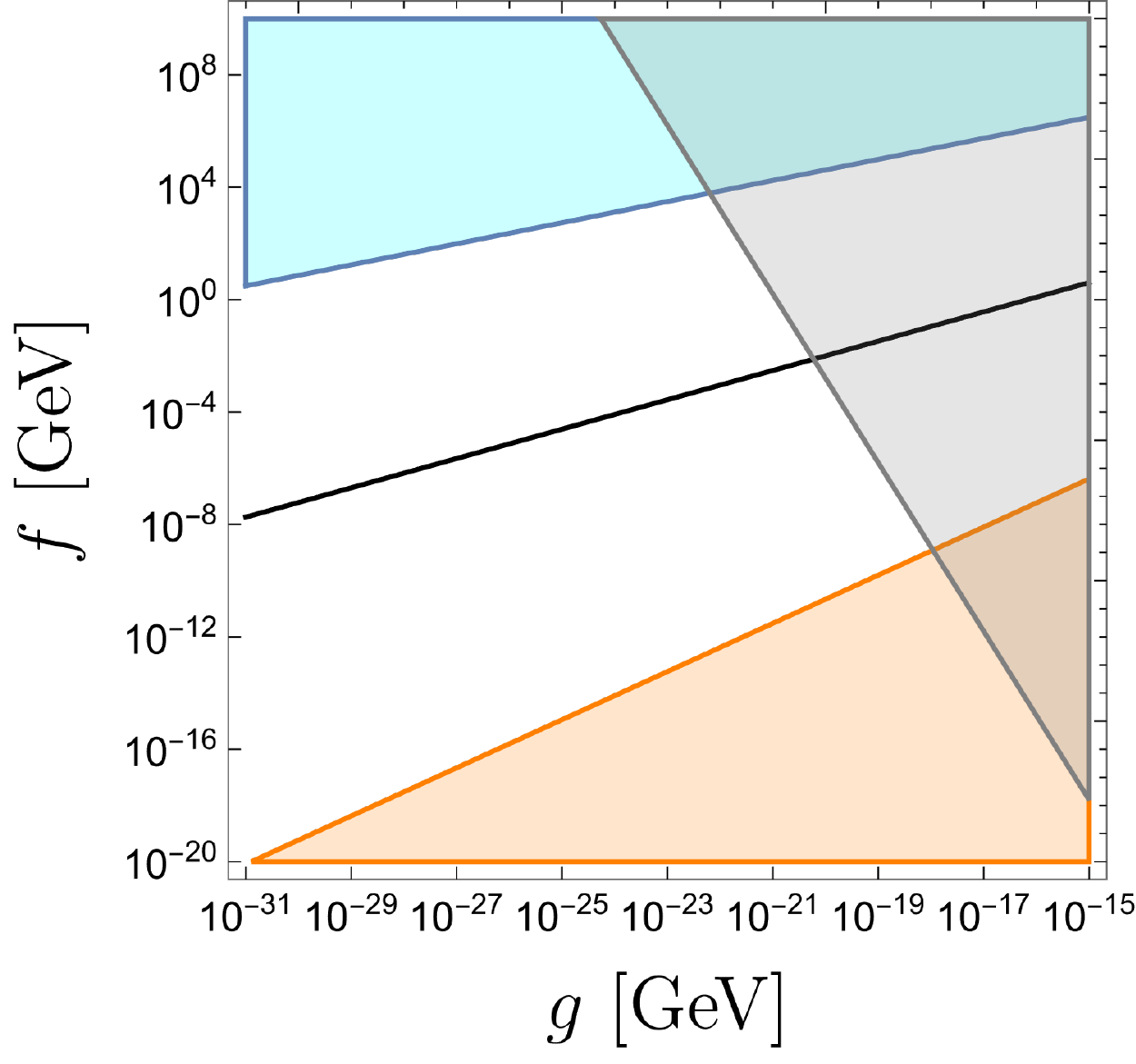}
    \caption{Viable parameter space: the colored regions are ruled out and the black line (c.f.~\eqref{Pzetaend}) yields the right amplitude of the scalar power spectrum $P_\zeta\sim 10^{-9}$. The gray region (c.f.~\eqref{solveCC}) 
    is ruled out as the CC would vary too quickly today. The orange region (c.f.~\eqref{faboveTend}) requires the CC relaxation to continue beyond the regime of validity of the higher dimensional operator that couples $\phi$ to the non-abelian gauge sector. The cyan region (c.f.~\eqref{strongwarm}) does not go through the strong regime of warm inflation. }
    \label{fig:crossplot}
\end{figure}

Our simple model achieves not only the tuning of the effective CC from a naturally large positive value to a small negative one, but also the right level of CMB anisotropies. The phenomenology of our model is largely determined by the slope $-g^3$ of the $\phi$ potential and the thermal-dissipation coefficient $\Upsilon$ describing the damping of $\phi$ during warm inflation in our model \cite{Berghaus:2019whh}
\begin{equation}
  \Upsilon=\frac{T^3}{f^2}  \label{upsilon}
\end{equation}
Here $f$ is related to the Lagrangian parameters $\alpha$ and $f_G$ through the relation $f\propto f_G/\alpha^{5/2}$ with an $O(0.1)$ proportionality constant that depends mildly on $\alpha$ and the details of the Yang-Mills sector. We will treat the more directly relevant quantity, $f$, together with $g$ as our free model parameters. The viable parameter space is shown in Fig.~\ref{fig:crossplot}, as we will show.

\section{Background \label{s:background}}
\subsection{Cold Inflation}
We suppose that the universe starts in a cold inflationary phase with a Hubble scale $H_i$, driven by the potential $V_{\rm i}$ of the slow-rolling $\phi$ field. In order to avoid eternal inflation, we require the classical rolling to dominate over quantum fluctuation, $\dot{\phi}_iH_i^{-1}\sim g^3/H_i^2\gtrsim H_i$, which can be rewritten as
\begin{equation}
    V_{\rm i}\lesssim M_{\rm P}^2g^2 \label{noeternal}
\end{equation}
where $M_{\rm P}$ is the reduced Planck mass. Hence, the highest possible initial CC, $V_{\rm i}$, that is amenable to our proposed CC-relaxation mechanism is set by the highest viable values of the slope parameter $g$ from the considerations below. As we will see, the combined requirements of solving the CC problem and having the dynamics of the CC relaxation be within the regime of validity of our effective field theory (EFT) puts an upper bound of $g\lesssim 10^{-18}\GeV$, corresponding to $V_{\rm i}\lesssim \GeV^4$. If we add to this the requirement of explaining the CMB anisotropies, $g$ is further limited to be $g\lesssim 10^{-20}\GeV$, which corresponds to $V_{\rm i}\lesssim \left(100\MeV\right)^4$.

\subsection{Thermalization of the Gauge Sector}
\label{ss:thermalizationofthegaugesector}

Meanwhile, the non-abelian gauge fields play no role as the universe inflates, that is, until they are considerably excited. The equations of motion for the gauge fields $A^{a\mu}$, after a spatial-Fourier and a helicity decomposition, reads
\begin{equation}
    \ddot{A}^{a\pm}_k+H\dot{A}^{a\pm}_k+\left(\omega_k^{\pm}\right)^2A^{a\pm}_k+(\text{non-abelian terms})=0 \label{Aeom}
\end{equation}
where $k\propto a^{-1}$ denotes a \textit{physical} wavenumber and 
\begin{equation}
   \left(\omega_k^{\pm}\right)^2=k\left(k\mp 2\xi H\right),\quad \xi\equiv \frac{\alpha\dot{\phi}}{4\pi f_GH} \label{xi}
\end{equation}
For a given $k$, the non-abelian terms in \eqref{Aeom} are negligible if $k\gtrsim e_G|\left<A^{a\mu}(t,x)\right>_{L\sim k^{-1}}|$, where $\left<A^{a\mu}(t,x)\right>_{L\sim k^{-1}}$ is the gauge-field amplitude in the configuration space averaged over an $L\sim k^{-1}$ sized domain. Initially, the gauge fields have only horizon-size fluctuations with the typical amplitude $\left|\left<A^{a\mu}(t,x)\right>\right|\sim H$, and so the evolution of modes with $k\gg e_GH$ is well described by the terms shown in \eqref{Aeom}, according to which $+$ helicity modes with $k<2\xi H$ are tachyonic. This provides a way to excite the gauge sector even if it begins in a Bunch-Davies vacuum.

The value of $\xi \propto H^{-2}$ increases as $\phi$ rolls down the potential with a Hubble-friction supported terminal velocity $\dot{\phi}=g^3/3H$. When $\xi$ goes considerably above unity, comoving gauge-field modes become tachyonic well before they cross the horizon and get amplified exponentially until the growth is stopped by Hubble friction slightly outside the horizon. The total energy density of the gauge fields soon reaches a steady state where the gain from tachyonic instability is balanced by Hubble dilution, at a value \cite{Anber:2009ua}
\begin{equation}
    -\frac{1}{4}\left<G^a_{\mu\nu}G^{a\mu\nu}\right>_H\sim N_c^2 10^{-4}\frac{e^{2\pi\xi}}{\xi^3} H^4 \label{GGH} 
\end{equation}
In the many Hubble times that follow, the steady-state gauge field energy density grows according to \eqref{GGH} as $\xi$ increases, until it reaches the value \cite{DeRocco:2021rzv}
\begin{equation}
    -\frac{1}{4}\left<G^a_{\mu\nu}G^{a\mu\nu}\right>_{\rm NL}\sim \frac{(\xi H)^4}{4\pi\alpha} \label{GGNL}
\end{equation}
whereupon non-abelian terms in \eqref{Aeom} become comparable to the abelian ones for $k\sim \xi H$ modes. By equating \eqref{GGH} and \eqref{GGNL}, one can calculate iteratively the $\xi$ when this occurs to be
\begin{equation}
    \xi_{\rm NL}\approx \left[\frac{6.7-\ln(N_c^2\alpha)}{2\pi}\right]+\frac{7}{2\pi}\ln\left[\frac{6.7-\ln(N_c^2\alpha)}{2\pi}\right]+\ldots \label{xiNL}
\end{equation}

At this point, some of the modes that would otherwise be growing tachyonically acquire positive non-abelian contributions to their squared frequency, terminating said growth. Non-linear interactions among these modes then lead to a net transfer of energy from the amplified modes to other less excited modes (including those with $-$ helicity) \cite{Kurkela:2011ti}. The entropy is maximized when the gauge sector is in thermal equilibrium with a peak energy $\sim T_{\rm th}$ that is given by energy conservation, i.e. by equating $\sim (N_c^2\pi^2/30)T_{\rm th}^4$ with the $-\left<G^a_{\mu\nu}G^{a\mu\nu}\right>/4$ at that time. However, such a thermal configuration
can only be achieved if the near-equilibrium thermalization rate $\sim 10N_c^2\alpha^2 T_{\rm th}$ is greater than $H$ \cite{Laine:2021ego} in the first place, which is equivalent to 
\begin{equation}
    -\frac{1}{4}\left<G^a_{\mu\nu}G^{a\mu\nu}\right>\gtrsim  3\times10^{-5}\frac{N_c^2}{\left(N_c\alpha\right)^8}H^4 \label{thrate}
\end{equation}
In what follows, we call the critical $\xi$ when the above is satisfied $\xi_{\rm scat}$.

We will focus on the case in which at the point the gauge fields hit the non-linear regime they have enough energy to immediately thermalize. This requires $\xi_{\rm NL}\gtrsim \xi_{\rm scat}$, which for $N_c={\cal O}(1)$ translates to
\begin{equation}
    N_c\alpha\gtrsim 0.1 \label{Nalpha}
\end{equation}
which comes from plugging \eqref{GGNL} and \eqref{xiNL} into \eqref{thrate}, and solving for $N_c\alpha$ numerically. The temperature $T_{\rm th}$ at which the gauge sector first thermalizes will be ${\cal O}(1)$ higher than the Hubble rate $H_{\rm th}$ at that time, and is roughly given by
\begin{equation}
    T_{\rm th}^2\sim H_{\rm th}^2\sim \frac{g^3}{f} \label{Hth}
\end{equation}
where we used $\dot{\phi}\sim g^3/3H$, \eqref{xi},  \eqref{xiNL}, and \eqref{Nalpha}. The opposite case, $\xi_{\rm NL}\lesssim \xi_{\rm scat}$, is also interesting. We choose to avoid it simply because we have less control of the evolution of the universe in that case, as it requires some knowledge of the non-linear dynamics of the non-abelian fields when $\xi\gtrsim \xi_{\rm NL}$ but before \eqref{thrate} is satisfied. We speculate on what would happen in that regime in Appendix~\ref{s:smallcouplingthermalization}.

In the above and in what follows, we assume that all the relevant scales of fluctuations (the higher of the Hubble rate $H$ or the temperature $T$ at a given time) during the CC relaxation are above the confinement scale of the gauge sector, i.e. the gauge sector is always weakly coupled. The lowest fluctuation energy scale occurs when the gauge sector begins to thermalize, when $H\sim H_{\rm th}$ (the Hubble rate $H$ was higher before and once the gauge sector thermalize its temperature is always higher than $H$ and will only increase). So, as long as $\alpha$ is perturbative at that point, it will also be perturbative at all other epochs.

\subsection{Warm Inflation}
Once the non-abelian gauge fields are thermalized, they behave as a thermal bath of dark radiation with energy density
\begin{equation}
    \rho_{\rm DR}=\frac{\pi^2}{30} g_* T^4 \label{radiation}
\end{equation}
This dark radiation then damps the motion of $\phi$ with a dissipative coefficient $\Upsilon=T^3/f^2$ as a form of backreaction. The number of degrees of freedom $g_*$ will depend on the gauge group of the Yang-Mills fields under consideration. For simplicity, we will from now on specialize to the $SU(2)$ case, for which $(\pi^2/30)g_*={\cal O}(1)$. 

The background equations in this epoch take the form
\begin{align}
    \ddot{\phi}+(3H+\Upsilon)\dot{\phi}-g^3&=0\label{bg1}\\
    \dot{\rho}_{\rm DR}&=\Upsilon \dot{\phi}^2-4H\rho_{\rm DR}\label{bg2}\\
    3M_{\rm P}^2H^2&=\rho_{\rm
    DR}+\frac{1}{2}\dot{\phi}^2-g^3\phi 
    \label{bg3}\\
    M_{\rm P}^2\dot{H}&=-\frac{1}{2}\dot{\phi}^2-\frac{2}{3}\rho_{\rm DR} \label{bg4}
\end{align}
These admit attractor solutions, dubbed warm inflation, where the universe is inflating ($V\gg \rho_{\rm DR}$ and $\dot{H}\ll H^2$) and thermal fluctuations dominate the fluctuations of $\phi$. During warm inflation, the rolling of $\phi$ is overdamped with negligible $\ddot{\phi}$, the thermal bath is kept in a quasi-equilibrium with negligible $\dot{\rho}_{\rm DR}$, and these imply 
\begin{equation}
    \dot{\phi}\approx \frac{g^3}{\Upsilon+3H},\quad \rho_{\rm DR}\approx \frac{\Upsilon}{4H}\dot{\phi}^2 \label{steadyphidotrad}
\end{equation}

In the parameter space regimes of our interest, the universe lands in the weak regime ($\Upsilon\lesssim H$) when the gauge sector first thermalizes at the temperature $T_{\rm th}$ found in \eqref{Hth}. Incidentally, this initial temperature is comparable to the the steady state value $T\sim g^6/f^2H_{\rm th}^3$, derived from \eqref{steadyphidotrad}, at that time. The radiation temperature then proceeds to track its steady-state value and scale as $T\propto H^{-3}$ until it is high enough to enter the strong regime ($\Upsilon\gtrsim H$). This happens when
\begin{equation}
    H_{\rm strong}\sim g \left(\frac{g}{f}\right)^{4/5}
\end{equation}
Once the universe is in the strong regime, the steady-state temperature scales more slowly with the Hubble rate, $T\propto H^{-1/7}$. It can be seen from the background equations that the conditions for slow-roll inflation, $V\gg \rho_{\rm DR}$ and $\dot{H}\ll H^2$, break at about the same time, when the Hubble rate and gauge-sector temperature are of the order of
\begin{equation}
    H_{\rm end}= \left(\frac{f^{4}g^{12}}{M_{\rm P}^7}\right)^{1/9},\quad T_{\rm end}=\left(M_{\rm P}f^2g^6\right)^{1/9}\label{Hend}
\end{equation}
marking the end of warm inflation. The field value at that point is around
\begin{equation}
    \phi_{\rm end}= -\left(\frac{f^8}{g^3M_{\rm P}^5}\right)^{1/9}M_{\rm P}
\end{equation}
 In order to have an extended period of strong warm inflation, where density perturbations are enhanced, we require $H_{\rm strong}$ to be well above $H_{\rm end}$, which amounts to
\begin{equation}
    f\ll \left(M_{\rm P}^5g^3\right)^{1/8} \label{strongwarm}
\end{equation}

\subsection{Crunching}

Though inflation has ended, $\phi$ continues to roll down at its thermal-friction-supported terminal velocity $\dot{\phi}_{\rm end}\approx g^3/\Upsilon_{\rm end}$. It soon crosses zero, overshoots to positive values, and passes a point where the now negative $V=-g^3\phi$ cancels out the $\rho_{\rm DR}\gg \dot{\phi}^2/2$ term in Eq.~\eqref{bg3} completely. At that point the universe comes to a halt ($H=0$) and begins to crunch ($H<0$). This Hubble sign flip causes the energy density $\rho_{\rm DR}$ of the dark radiation to start blueshifting as $a^{-4}$. Soon, it takes over as the dominant component and remains to be so for most of the crunching phase. Due to the increasing thermal friction $\Upsilon\propto T^3$ and Hubble rate $H\propto T^2$ during the crunching, the distances $\phi$ travels in the subsequent Hubble times continue to diminish. Hence, most of the $\phi$ displacement after the end of warm inflation is already covered in the first Hubble time or so, and $\phi$ eventually parks at a value $\sim \dot{\phi}_{\rm end}H_{\rm end}^{-1}\sim \left|\phi_{\rm end}\right|$.

\begin{figure}
    \centering
    \includegraphics[width = \linewidth]{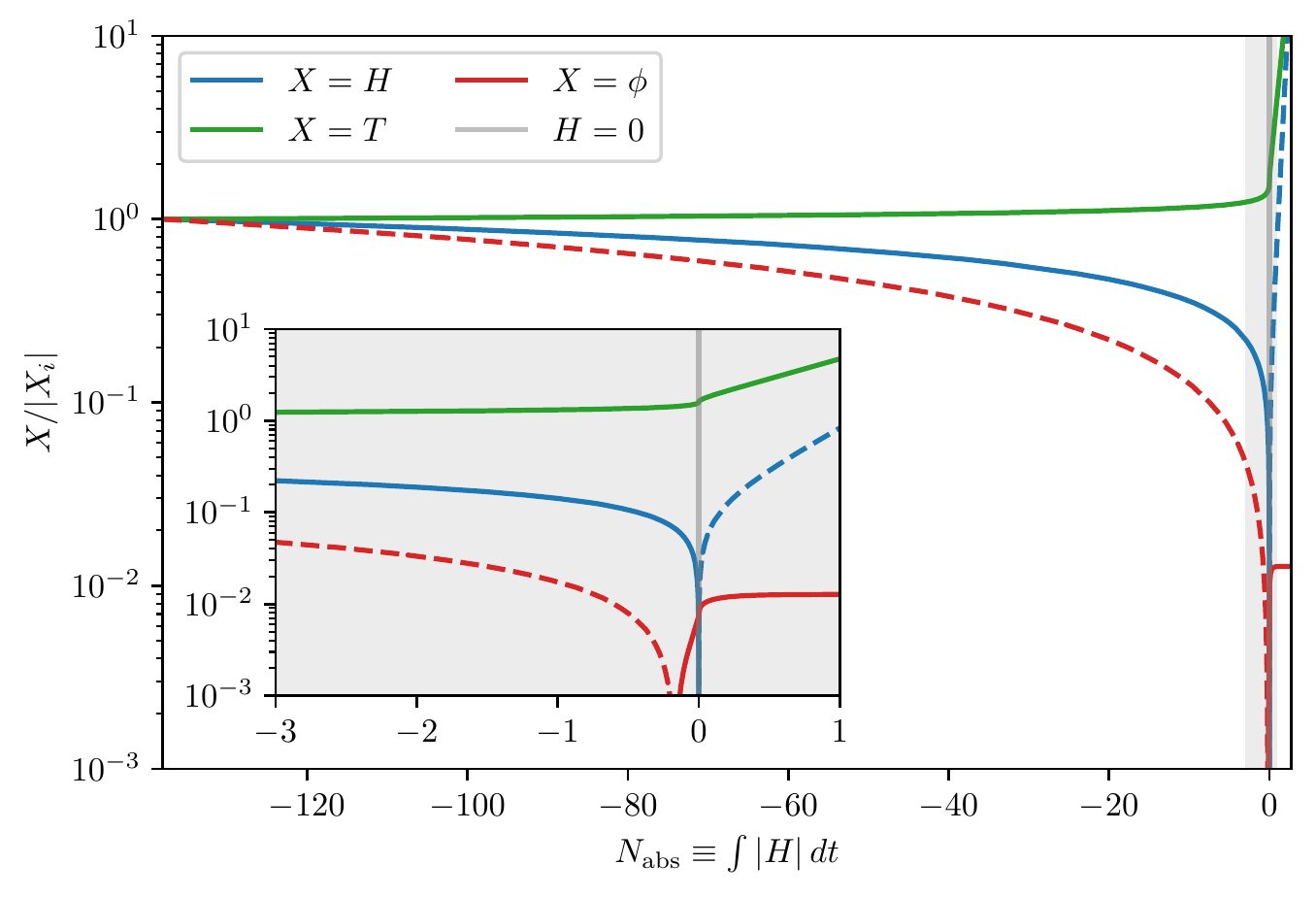}
    \caption{Evolution of $\phi$, $T$, and $H$ from the slow-roll regime to where $\phi$ stalls in the crunching stage before the bounce. Dashed lines indicate negative values. Quantities are normalized to their initial slow-roll value, and are plotted against the absolute number of $e$-folds. The inset plot is a zoom-in around the $H=0$ point. Here, we take $f=100\,{\rm keV}$ and $g=10^{-23}\,{\rm GeV}$.}
    \label{fig:fiducial}
\end{figure}

We performed a numerical simulation to confirm the background evolution around the end of warm inflation and the start of crunching discussed above (see Fig.~\ref{fig:fiducial}). We set $f=100\,{\rm keV}$, $g=10^{-23}\,{\rm GeV}$, and evolve the equations of motion numerically using the absolute number of $e$-folds $N_{\rm abs}\equiv\int\left|H\right| dt$ relative to the $H=0$ point when the universe starts crunching as a proxy for time. The results suggest that the Hubble rate does not change significantly in the last tens of $e$-folds of the warm inflation. The same plot also shows that the stalling value of $\phi$ is comparable to its absolute value $\left|\phi_{\rm end}\right|$ at the end of warm inflation (a few $e$-folds before the $H=0$ point).

\subsection{Bouncing and Expanding to the Present Epoch}
At this point, the universe is still crunching, the CC is tuned to a tiny but negative value, and the Standard Model sector is empty. The details of what happens next will depend on the full high energy theory that takes over outside the realm of validity of the EFT shown in Eq.~\eqref{model}. That said, we know that several things need to happen in order to produce the observed features of the universe: (1) the universe must bounce (stop crunching and start expanding), (2) the Standard Model sector must be excited, and (3) the CC must be made positive and at the right level. The bounce can be achieved either by violating NEC \cite{Alberte:2016izw} or by inducing vorticity \cite{Graham:2017hfr}. It is conceivable that the Standard Model sector as well as the degrees of freedom responsible for the bounce, whatever the mechanism is, are generated as the radiation is blue-shifted to sufficiently high temperatures. While we expect the $\phi$ contribution to the CC, $-g^3\phi$, to remain negative, the CC also receives contributions from other sectors that are decoupled from the relaxation dynamics of $\phi$. Some of these sectors may have multiple vacua and possibly transitioned between them as the universe was heated before the bounce and cooled after the bounce\footnote{A concrete realization of this scenario can be found in Appendix~A of \cite{Graham:2019bfu}.}. Such a vacuum switching adds a $\phi$-independent contribution to the CC which would show up in the EFT under consideration as a constant shift $\Delta V$ in the potential $V$ of $\phi$
\begin{equation}
    V=-g^3\phi\quad\longrightarrow\quad  V=-g^3\phi+\Delta V
\end{equation}
This shift must be positive in order to explain the observed CC value today.

In the interest of solving the CC problem, we require our relaxation process to tune the CC to a negative value whose magnitude $\sim g^3\left|\phi_{\rm end}\right|$ is comparable to or less than the shift $\Delta V$ which, in turn, must be comparable or close to the value observed today $\sim (2\meV)^4$,
\begin{equation}
    g^3\left|\phi_{\rm end}\right|\lesssim \Delta V\sim  (2\meV)^4 \label{CCend}
\end{equation}
This theoretical constraint allows the process that shifts the relaxed tiny negative CC to the positive value observed today to be natural, i.e. not fine-tuned. Apart from that, we also want to avoid the possibility that the now correct CC, $V\sim (2\meV)^4$, would un-tune itself due to the subsequent rolling of $\phi$. Whether, or to which degree, this occurs has some dependence on the UV completion of our EFT (Eq.~\eqref{model}). We assume here that the thermal friction on $\phi$ increases with temperature, in which case most of the $\phi$ rolling would occur at low energies where our EFT is valid,
and leave the discussion on the opposite possibility (that the thermal friction diminishes at high temperatures) to the next subsection.

The most relevant constraint on the dark energy today comes from its measured energy density $\sim (2\meV)^4$ and the constraints on its equation of state $w_{\rm DE}$, which in our model is given by
\begin{equation}
    w_{\rm DE}\approx\frac{-(2\meV)^4+\dot{\phi}_0^2/2+\rho_{\rm DR,0}/3}{(2\meV)^4+\dot{\phi}_0^2/2+\rho_{\rm DR,0}}
\end{equation}
The Planck TT, TE, EE+lowE+lensing+Sne+BAO constraints the equation of state today to be $-1<w_{\rm DE}<-0.95$ \cite{Planck:2018vyg}. Note that in cases of our interest $\rho_{\rm DR}\gg (\Upsilon/H)\dot{\phi}^2\gg \dot{\phi}^2$, and so the latter constraint is essentially a constraint on the amount of dark radiation today:
\begin{equation}
    \rho_{\rm DR,0}\lesssim 0.6 \meV^4
\end{equation}
In the likely scenario where $\dot{\phi}$ and $\rho_{\rm DR}$ are tracking the steady state solution \eqref{steadyphidotrad} today\footnote{Assuming that the $\rho_{\rm DR}$ at the time of BBN is consistent with the $\Delta N_{\rm eff}$ bound, $\dot{\phi}$ and $\rho_{\rm DR}$ would naturally evolve to reach the steady state given by \eqref{steadyphidotrad} before today.}, we can use \eqref{steadyphidotrad} to express $\rho_{\rm DR}$ in terms of the current Hubble rate $H_0\approx 10^{-42}\GeV $ and the model parameters $f$ and $g$
\begin{equation}
    \rho_{\rm DR,0}\sim \left(\frac{f^2g^6}{4H_0}\right)^{4/7}\lesssim 0.6\meV^4 \label{solveCC}
\end{equation}
It turns out that this is  more stringent than the theoretical constraint \eqref{CCend}, and so can be regarded as our model requirement for solving the cosmological constant problem.

\subsection{EFT Validity}
The EFT under consideration, Eq.~\eqref{model}, is trustworthy throughout the CC-relaxation process if at any given time the cutoff\footnote{We adopt $f_G/\alpha$ as the cutoff of our EFT here, but the actual cutoff scale will depend on how the EFT is UV completed. For example, in some UV completions with $\phi$ coupled to heavy fermions \cite{Kim:1979if, Shifman:1979if}, the scale at which the EFT breaks down is closer to $f_G$ than $f_G/\alpha$.} $f_G/\alpha\sim 10 \alpha^{3/2}f$ of the higher dimensional operator that couples $\phi$ to the Yang-Mills sector is above the highest relevant fluctuation scale, i.e. the higher of the Hubble rate and the gauge sector temperature. The Hubble rate is the only relevant scale during cold inflation and its highest value $H_i$ occurs at the very beginning, while the temperature $T$ dominates over $H$ throughout warm inflation and is highest at the end of it. In cases of our interest, we always have $T_{\rm end}\gg g\gtrsim H_i$. Thus, it suffices to require 
\begin{equation}
    \alpha\gtrsim \left(\frac{T_{\rm end}}{f}\right)^{2/3}
    \label{faboveTend}
\end{equation}
Since Section~\ref{ss:thermalizationofthegaugesector} we have specialized to the regime where $0.1 \lesssim \alpha\lesssim 1$ (for  $N_c={\cal O}(1)$) at the time of the gauge sector thermalization. The running of the coupling constant for a pure Yang-Mills theory tells us that by the end of warm inflation, we would have $\alpha\sim \left[1/ 0.1+(11/6\pi)N_c\ln\left(T_{\rm end}/H_{\rm th}\right)\right]^{-1}\sim 0.01$. In light of this, \eqref{faboveTend} turns into a constraint on $f$ and $g$.

Let us return to the issue of the CC un-tuning itself due to the rolling of $\phi$. We have discussed previously the case where the thermal friction on $\phi$ becomes stronger with increasing temperature in the UV regime, in which case most of the $\phi$ rolling occurs at low energies. We now consider the extreme scenario where the thermal friction disappears completely outside of the realm of validity of our EFT, assuming that the UV regime is radiation dominated throughout. In that case, when the temperature goes above the cutoff $f_G/\alpha$ during the contraction, at the Hubble scale $H_{\rm cutoff}\sim 10^2\alpha^3 f^2/M_{\rm P}$, the initially small $\dot{\phi}$ (due to thermal friction) immediately accelerates to $g^3/H_{\rm cutoff}$ in a Hubble time due to the slope $-g^3$ of the $\phi$ potential. Following that, $\dot{\phi}$ scales up as $a^{-3}$ due to Hubble anti-friction up to the bounce point and proceeds to scale down as $\dot{\phi}\sim a^{-3}$ (all the while with $\dot{\phi}\gtrsim g^3/H$) until $\dot{\phi}$ is back at $g^3/H_{\rm cutoff}$ at $H\sim H_{\rm cutoff}$, where the thermal friction turns back on and slows $\dot{\phi}$ down significantly. Given this scaling and that $H^{-1}\propto a^2$ during radiation domination, the field excursion per Hubble time $\dot{\phi}H^{-1}$ would be maximum at the bounce point, where $\dot{\phi}H^{-1}\sim \left(g^3/H_{\rm cutoff}^2\right)\left(T_{\rm bounce}/10\alpha^{3/2}f\right)$. Requiring the CC to not change by more than $\sim (2\meV)^4$ at that time amounts to
\begin{equation}
    \frac{T_{\rm bounce}}{10^{16}\GeV}\lesssim \left(\frac{\alpha}{0.01}\right)^{15/2}\left(\frac{f}{10\MeV}\right)^5\left(\frac{g}{10^{-20}\GeV}\right)^{-6}
\end{equation}
To reiterate, this is only when the thermal friction disappears as soon as the temperature is above the cutoff. Thus, this is a model-dependent constraint. Since this is not a forbidding constraint, the success of the CC relaxation per se does not hinge on the UV physics, even if the only viable UV model requires the friction to disappear.

\section{Perturbations \label{s:perturbations}}
CMB observations by the Planck satellite found at 95\% confidence level the following amplitude and spectral index of the scalar power spectrum  \cite{Akrami:2018odb}
\begin{equation}
    P_\zeta(k_*)\approx 2.1\times 10^{-9},\quad n_s(k_*)=0.9649\pm 0.0084 \label{CMBobs}
\end{equation}
at around the pivot scale whose wavenumber today is $k_*=0.05\text{ Mpc}^{-1}$. These observed perturbations are often attributed to the vacuum fluctuations during an inflationary period taking place at some energy scale well above MeV, the big bang nucleosynthesis (BBN) temperature, in a universe that has a beginning and has been expanding since.

Our CC-relaxation model does include a period of inflation, albeit one that is separated by a crunch and a bounce from the hot big bang epoch. When it comes to explaining CMB anisotropies, this fact by itself is not an issue as the CMB scale perturbations are frozen as soon as they turn superhorizon during the pre-bounce inflation and therefore insensitive to the subsequent cosmological history until around the epoch of recombination \cite{Creminelli:2004jg}. A robust feature of our CC-relaxation model is that the last stages of the pre-bounce inflation must occur at $\lesssim \meV$ energy scales in order to solve the cosmological constant problem. Given the fact that the vacuum contribution to the scalar power spectrum $P_\zeta$ is related to the inflationary Hubble rate $H$ as $P_\zeta\propto H^2$, its low energy scale poses a significant challenge in getting the right level of curvature perturbations.

We show in this section that thermal perturbations during warm inflation can provide enough enhancements of the curvature perturbations relative to the vacuum one to match the amplitude seen in CMB. However, due to the lack of curvature in the linear potential $-g^3\phi$ in our model, the resulting scalar power spectrum is always blue-tilted, in contrast to the red-tilted spectrum suggested by CMB observations. This is, nevertheless, a mild problem that can be addressed in isolation, while respecting the required shift symmetry for the CC relaxation.

Moreover, our pre-bounce inflationary period automatically solves all the other problems that the standard, post-big-bang inflation solves. The horizon and flatness problems are solved by the sheer amount of expansion during the inflationary period as in traditional inflationary scenarios. While the thermal fluctuations during warm inflation give rise to highly amplified scalar perturbations, they do not source extra tensor perturbations. The tensor perturbations remain to be sourced by quantum fluctuations, which in our case is suppressed due to the smallness of the Hubble rate near the end of warm inflation. Hence, our model predicts a vanishingly small tensor to scalar ratio $r$. Warm inflation also yields unique non-gaussianity signatures that are within the current observational bounds. In the strong dissipative regime that is of interest to us, $\Upsilon\gg 3H$, warm inflation predicts $f_{\rm NL}={\cal O}(10)$ \cite{Moss:2011qc,Berghaus:2019whh}.

\subsection{Scalar Amplitude}

The curvature power spectrum for modes exiting the horizon during the strong warm inflation (with $\Upsilon\gg H$) is given by \cite{Bastero-Gil:2011rva}
\begin{equation}
    P_\zeta\approx 5\times 10^{-10}\left(\frac{\Upsilon}{3H}\right)^{8}\left(\frac{T}{H}\right)\left(\frac{H^2}{\dot{\phi}}\right)^2 \label{scalarpowerspectrum}
\end{equation}
The exact amplitude $P_\zeta(k_*)$ evaluated at the CMB pivot scale down to ${\cal O}(1)$ factors will depend not only on the spectral index but also on the number of $e$-folds separating the horizon exit of the pivot scale and the end of warm inflation. These are more model-dependent and will be discussed in the next subsection. Our focus here is on getting the amplitude at the right order of magnitude. An extrapolation of the small spectral tilt observed in the CMB suggests that $P_\zeta(k)$ would not vary by orders of magnitudes in the range of scales that become superhorizon in the last tens of $e$-folds of warm inflation. Hence, a good estimate for $P_\zeta(k_*)$ is given by Eq.~\eqref{scalarpowerspectrum} for the mode that exited the horizon at end of warm inflation
\begin{equation}
    P_\zeta\sim 10^{-13}\left(\frac{g^{12}M_{\rm P}^{11}}{f^{23}}\right)^{2/3} \label{Pzetaend}
\end{equation}
where we have used \eqref{upsilon}, \eqref{steadyphidotrad}, \eqref{Hend}, and taken $T\sim T_{\rm end}$, etc. Thus, $P_\zeta(k_*)\sim 10^{-9}$ can be achieved with
\begin{equation}
    f\sim g^{12/23}M_{\rm P}^{11/23} \label{fscalaramplitude}
\end{equation}
Given the strong dependence of the amplitude Eq.~\eqref{Pzetaend} on $f$ and $g$, we expect this rough parameteric requirement to be robust against model variations. To explain both the CMB anisotropies and the CC problem at once, we need to satisfy the above plus the dark energy constraint Eq.~\eqref{solveCC}. This amounts to
\begin{equation}
    g\lesssim  10^{-20}\GeV,\quad f\lesssim 10\MeV\label{gCCprob}
\end{equation}

\subsection{Scalar Spectral Index}
\label{ss:scalarspectralindex}
The spectral index evaluated at the pivot scale $k_*$ is given by
\begin{align}
    n_s-1=\left.\frac{d\ln P_\zeta}{dN}\right|_{k=k_*}=\frac{6}{7}\left(11\epsilon_V-8\eta_V\right) \label{spectralindex}
\end{align}
where $dN$ is differential number of $e$-folds and
\begin{eqnarray}
    \epsilon_V &\equiv& \frac{M_{\rm P}^2}{2(1+\Upsilon_*/3H_*)}\left(\frac{V_*'}{V_*}\right)^2 \nonumber\\ \eta_V &\equiv& \frac{M_{\rm P}^2}{1+\Upsilon_*/3H_*}\left(\frac{V_*''}{V_*}\right)
\end{eqnarray}
are two of the warm-inflation slow-roll parameters evaluated at the horizon crossing of the pivot scale. In the minimal model where the potential is perfectly linear, $V(\phi)=-g^3\phi$, and $\phi$ interacts only with the weakly-coupled non-abelian gauge sector that provides the dominant friction during warm inflation, we have $\eta_V=0$ and consequently the scalar power spectrum is always blue-tilted $n_s-1>0$.

\begin{figure}
    \centering
    \includegraphics[width=\linewidth]{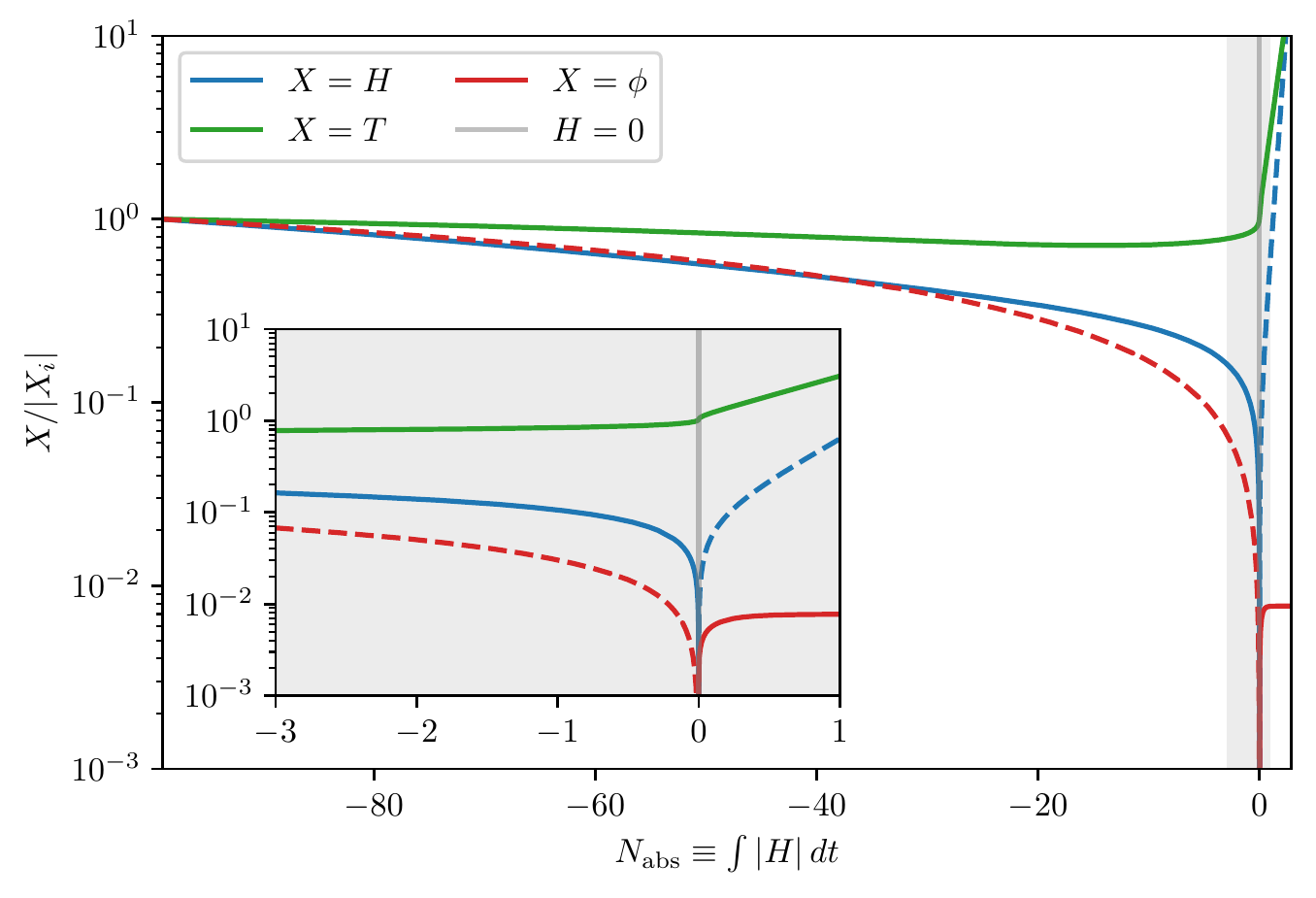}
    \includegraphics[width=\linewidth]{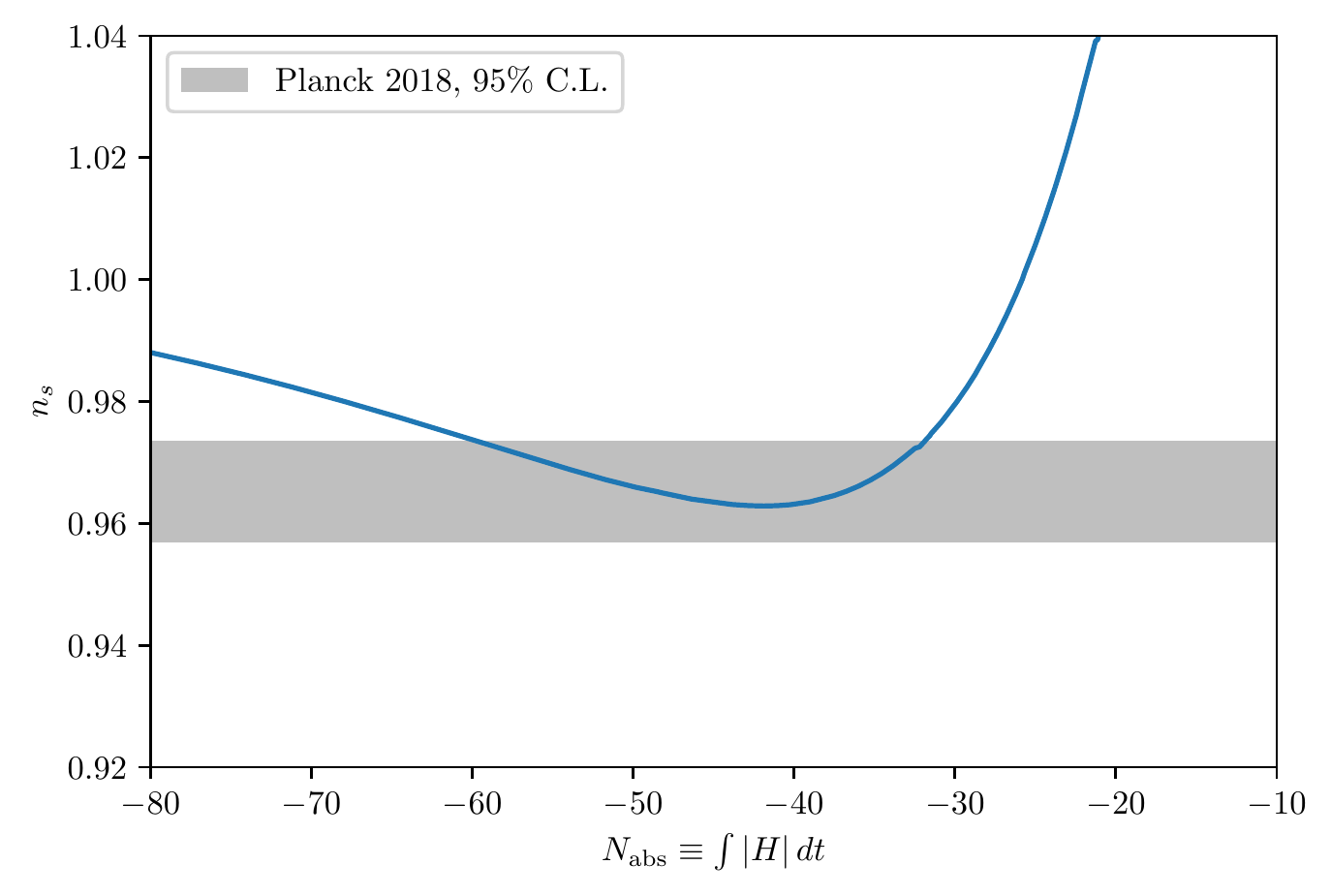}
    \caption{The top panel here is the same as Fig.~\ref{fig:fiducial}, but for the modulated potential Eq.~\eqref{eqn:modulated-potential}. It shows that the background evolution remains unchanged qualitatively. The bottom panel shows the scalar spectral index $n_s$ as a function of the absolute number of $e$-folds, with the gray band showing the Planck constraints. Here, we take $f=100\,{\rm keV}$, $g=10^{-23}\,{\rm GeV}$, $\tilde f=2 \times 10^{14}\,{\rm GeV}$, $\phi_0 = 4.5 \tilde f$, and $\tilde \Lambda = 0.97 (\tilde f g^3)^{1/4}$.}
    \label{fig:tweak}
\end{figure}

To get a red-tilted spectrum, we need a non-zero and sufficiently positive $\eta_V$ to counter the $\epsilon_V$ term in Eq.~\eqref{spectralindex}. This can be achieved, for example, by coupling $\phi$ to a cold, confined non-abelian sector with a decay constant $\tilde{f}$ and a confinement scale $\tilde{\Lambda}$. Doing so adds a modulation to the linear potential $-g^3\phi$
\begin{equation}\label{eqn:modulated-potential}
    V(\phi)=-g^3\phi+\tilde{\Lambda}^4\left[\cos\frac{\phi_0}{\tilde{f}}-\cos\left(\frac{\phi-\phi_0}{\tilde{f}}\right)\right]
\end{equation}
The phase $\phi_0$ accounts for the separation between the zero of $V(\phi)$ that is relevant to our scenario and the nearest extremum of the potential modulation. The constant shift $\tilde{\Lambda}^4\cos(\phi_0/\tilde{f})$ ensures that the potential still crosses zero when $\phi$ does. We require $\tilde{\Lambda}^4/\tilde{f}\leq g^3$, so that the resulting potential alternates between steep cliffs and flat plateaus with no minima. Otherwise $\phi$ could be trapped prematurely in one of these minima. This no-trapping condition and the definition of $\eta_V$ amount to an upper bound on $\tilde{f}$, which can be written as
\begin{equation}
    \tilde{f}<\frac{\cos\left[(\phi_*-\phi_0)/\tilde{f}\right]}{\eta_V}\left(\frac{g^3}{V'_*}\right)\dot{\phi}_*H_*^{-1}
\end{equation}
assuming that the CMB scales became superhorizon before slow-roll breaks. It is likely that the CMB scales $k\sim 10^{-4}-10^{-1}\text{ Mpc}^{-1}$, including the pivot scale $k_*$, all exited the horizon on one of the plateaus since the rolling of $\phi$ is significantly slower there, and it must be on a part of the plateau where $V''>0$ so that the resulting spectral index is red. This would require the field excursion over the $\Delta N_{\rm CMB}$ $e$-folds during which the CMB scales exited the horizon to be much less than the oscillation period $2\pi \tilde{f}$ of the potential, which sets a lower bound on $\tilde{f}$
\begin{equation}
    \tilde{f}\gg\frac{\Delta N_{\rm CMB}}{2\pi}\dot{\phi}_*H_*^{-1}
\end{equation}
Given that $\Delta N_{\rm CMB}\sim 10$ and $\eta_V\sim 1\%$, the following choice of parameters 
\begin{equation}
    \tilde{f}\sim 10^2\dot{\phi}_*H_*^{-1},\quad \tilde{\Lambda}^4\sim \tilde{f}g^3
\end{equation}
meet the requirements discussed above (with a little tuning). For example, if $f\sim 100\keV$ and $g\sim 10^{-23}\GeV$ then $\tilde{f}\sim 10^{14}\GeV$ and $\tilde{\Lambda}\sim 10^{-14}\GeV$ would work. Figure~\ref{fig:tweak} shows that, for appropriate choices of $\tilde{f}$ and $\tilde{\Lambda}$, there is a range of $e$-folds $\Delta N\gtrsim 10$ with $n_s$ lying inside the CMB 95\% confidence level band, Eq.~\eqref{CMBobs}.  We have also checked that the running of the spectral index is well within the the 95\% confidence level limits from Planck. Note that modes continue to exit the horizon in the radiation-dominated contraction following the end of warm inflation. To account for this, we have chosen the parameters such that the pivot scale crossed the horizon at around $N\approx 40$ $e$-folds before the end of warm inflation, instead of the typically assumed $N\approx 60$.

\section{Discussion and Conclusion \label{s:discussion}}

We have developed a model that dynamically relaxes the (effective) cosmological constant to a tiny value through the rolling of a shift-symmetric scalar field down its linear potential. This simple model not only explains the unexpected smallness of the cosmological constant and the reheating of the universe to the hot big bang era, but also gives rise to the right amplitude of the primordial density perturbations at the CMB scales. It also improves the degree of cosmological constant tuning and is even simpler compared to the previously proposed model of the same kind \cite{Graham:2019bfu}. Remarkably, this class of models is already strongly constrained by the shift symmetry required to ensure the radiative stability of the $\phi$ field, and yet our model can give rise to a rich cosmology that achieves the above-mentioned feats in a way that is consistent with all the current observations. It will be interesting to attempt a more systematic and generalized approach to realizing this class of solutions. Possible extensions include considering different dissipative mechanisms and multiple rolling scalars.

A number of assumptions were made in reaching our conclusions. We assumed that the non-linear self-interactions of the gauge fields tend to bring themselves to a thermal equilibrium, but did not describe the thermalization process in full. Numerical simulations will provide a fuller picture of the thermalization process in the nonlinear regime and a dedicated study may reveal analytically-calculable processes that lead to the thermalization of the gauge fields, possibly before they turn nonlinear. The simple EFT shown in \eqref{model} that achieves the CC relaxation and generates the right amplitude of primordial density perturbations does not on its own provide: a concrete realization of the non-singular bounce, a mechanism that turns the relaxed tiny negative CC into the positive value $(2\meV)^4$ observed today, and an explanation for the slight red tilt of the scalar power spectrum seen in the CMB. While we have provided examples for each of these, it would be interesting to write down a complete and renormalizable model that achieves all the key steps in this scenario, as it would provide further consistency checks.

The rich cosmology of our scenario gives rise to a series of predictions that can be checked against current and future cosmological surveys. First, the range of scales spanning 60 or so $e$-folds that re-enter the horizon during the post-bounce expansion will include not only the modes that turn superhorizon during warm inflation, but also those during the subsequent contraction period. The resulting scalar power spectrum will have a mixed feature where it is nearly scale-invariant at large scales (the inflation part) and perhaps strongly blue-tilted at small scales (the contraction part), with a rapid transition somewhere in between. Furthermore, the non-gaussianities in the primordial perturbations generated during the warm inflation period in our model are small enough to be within the current constraints, but can be tested in future observations. Our model also predicts highly suppressed inflationary gravitational waves, however the cosmological bounce that occurs afterward may generate potentially observable gravitational waves with a peaked spectrum.

Unlike most early universe models, the most defining parts of our model (i.e. those described by the EFT shown in Eq.~\eqref{model}) operates at low energies, and this gives us further opportunities to test it today. If it is indeed realized in nature, the present-day cosmology should mirror to some extent what happened in the pre-bounce, low-energy universe. In fact, our model says that we are about to enter a new period of warm inflation. Most directly related to the dynamical explanation for the cosmological constant problem is the evolving nature of the dark energy. This can be probed through the time variation of its equation of state today. The strong friction needed to generate the right level of primordial density perturbations requires a thermal bath of dark radiation coupled to the dark energy. Assuming even weak mixing with the standard model, such a dark radiation could be probed directly in a variety of ways \cite{Berghaus:2020ekh}.

\section*{Acknowledgements}
We thank William DeRocco, Michael Fedderke, Peter Graham, and Saarik Kalia for useful discussions. DK and SR are supported in part by the NSF under grant PHY-1818899.   SR is also supported by the DoE under a QuantISED grant for MAGIS and the SQMS quantum center and the Simons Investigator award 827042.

\newpage
\appendix
\section{Gauge Sector Thermalization in the Small Coupling \boldmath $\alpha$ Regime} \label{s:smallcouplingthermalization}

As argued in Section~\ref{ss:thermalizationofthegaugesector}, one requirement for the gauge sector to thermalize is that it must have enough energy to satisfy \eqref{thrate}. If $\xi_{\rm NL}\lesssim \xi_{\rm scat}$, the gauge sector does not immediately thermalize when it becomes non-linear, and will stay in the non-linear regime without thermalizing until it acquires enough energy from $\phi$. In this appendix, we argue that the gauge sector would still thermalize in this case and estimate the possible values of the Hubble $H_{\rm th}$ rate when it does. The challenge lies mainly in determining the growth rate of the energy density of the gauge fields in the non-linear regime, where \eqref{GGH} no longer applies. The gauge field temperature right after thermalization is then given by $T_{\rm th}\sim H_{\rm th}/10N_c^2\alpha^2$, which can be much higher than the Hubble rate at that time, $H_{\rm th}$, depending on how small $N_c\alpha$ is.

Let us first assume that when $\xi\gtrsim \xi_{\rm NL}$, the universe settles to a steady state where the kinetic energy of $\phi$ is approximately constant in a Hubble time:
\begin{equation}
    \frac{d}{dt}\left(\frac{\dot{\phi}^2}{2}\right)=g^3\dot{\phi}-3H\dot{\phi}^2-Q\approx 0 \label{assumesteady}
\end{equation}
where
\begin{equation}
    Q=-\frac{1}{4}\left<G^a_{\mu\nu}G^{a\mu\nu}\right>_{\rm NL}\Gamma\sim \frac{(\xi H)^4}{4\pi \alpha}\Gamma
\end{equation}
is the energy density transfer rate from $\phi$ to the gauge sector, parameterized by the rate $\Gamma$ to be estimated later. In order to thermalize, the energy density acquired by the gauge sector per Hubble time $QH^{-1}$ must satisfy \eqref{thrate}, which for $N_c={\cal O}(1)$ happens when
\begin{equation}
    \xi_{\rm th} \sim \left(\frac{H}{\Gamma}\right)^{1/4}\frac{1}{\alpha^{7/4}}\label{xithlowalpha}
\end{equation}
From \eqref{assumesteady}, we see that there are two possible steady states for $\dot{\phi}$ at that time
\begin{equation}
    \dot{\phi}_{\rm th}\sim \text{min}\left[\frac{g^3}{H_{\rm th}}, \frac{Q}{g^3}\right]\sim \text{min}\left[\frac{g^3}{H_{\rm th}}, \frac{H_{\rm th}^5}{\alpha^8 g^3}\right] \label{philowalpha}
\end{equation}
depending on how small $\alpha$ is. The Hubble rate at which the gauge sector thermalizes $H_{\rm th}$ can then be found from \eqref{xithlowalpha}, \eqref{philowalpha}, and the definition of $\xi$ (Eq.~\eqref{xi})
\begin{equation}
    H_{\rm th}\sim \text{max}\left[\xi_{\rm th}^{-\frac{1}{2}}\alpha^{-\frac{3}{4}}\left(\frac{g^3}{f}\right)^{\frac{1}{2}},\xi_{\rm th}^{\frac{1}{4}}\alpha^{\frac{19}{8}}(g^3f)^{\frac{1}{4}}\right] \label{Hlowalpha}
\end{equation}

We now attempt to estimate $\Gamma$. The thermalization condition \eqref{thrate} requires the energy density in the gauge sector to go well above \eqref{GGNL}. If at that time the energy density is still concentrated around the tachyonic $k\sim \xi H$ modes, we would have $e_G|\left<A^{a\mu}(t,x)\right>_{L\sim (\xi H)^{-1}}|\gg e_G^{1/2}\left(-\left<G^a_{\mu\nu}G^{a\mu\nu}\right>_{\rm NL}/4\right)^{1/4}\sim \xi H$. It follows that $k\lesssim e_G|\left<A^{a\mu}(t,x)\right>_{L\sim k^{-1}}|$ for most of these modes and this shuts off their tachyonic instabilities. While the tendency to maximize entropy implies that energy would continue to flow from $\phi$ to the gauge sector even when the latter has become completely nonlinear and no longer tachyonic, it is possible that the energy transfer rate in that regime is highly suppressed. To be on the conservative side, we assume that tachyonic instability is the only way for the gauge fields to acquire energy from $\phi$. The tachyonic band would re-open if the configuration-space gauge-field amplitude is somehow reduced. There are two ways for this to happen. First, Hubble friction will eventually deplete it if the gauge sector stops gaining energy from $\phi$. The second way is through the self-scatterings among the nonlinear modes, which tend to redistribute energy from the amplified low-wavenumber modes to the less-occupied high-wavenumber modes. This has the effect of increasing the typical gradient of $A^{a\mu}$. The average amplitude on the tachyonic scales  $|\left<A^{a\mu}(t,x)\right>_{L\sim (\xi H)^{-1}}|$ must therefore go down due to cancellations (the same conclusion can be reached from energy conservation considerations). Based on these considerations, we expect $\Gamma$ to lie in the range $H\lesssim \Gamma\lesssim \xi H$. The lowest $\Gamma$ corresponds to the case where the self-scattering rates are so slow that the gauge fields need to wait a Hubble time before the tachyonic band opens up again due to Hubble dilution, while the highest $\Gamma$ is expected when the self-scattering rates are so fast that the tachyonic band is essentially always open. It follows from \eqref{xithlowalpha} that $\alpha^{-7/4}\lesssim \xi_{\rm th}\lesssim\alpha^{-7/5}$ and this can be plugged into \eqref{Hlowalpha} to give the Hubble rate when the gauge fields thermalize.

\newpage
\bibliography{references}
\bibliographystyle{h-physrev}

\end{document}